\documentclass[aps,pre,showpacs,floatfix,twocolumn,superscriptaddress,10pt]{revtex4-1}

\usepackage[T1]{fontenc}

\usepackage{bm}				
\usepackage{amsmath}
\usepackage{amssymb}
\usepackage{latexsym}
\usepackage{amsfonts}

\usepackage{graphicx}

\newcommand{\ket}[1]{\vert{#1}\rangle} 
\newcommand{\bra}[1]{\langle{#1}\vert} 
 
\newcommand{\proj}[1]{\ket{#1}\!\bra{#1}}
\newcommand{\mean}[1]{\langle #1 \rangle}
\newcommand{\one}{\openone}
\DeclareMathOperator{\Tr}{Tr}

\newcommand{\abs}[1]{\left|#1\right|} 

\newcommand{\beq}{\begin{equation}}
\newcommand{\eeq}{\end{equation}}

\renewcommand{\vec}[1]{\boldsymbol{#1}}

\begin{document}

\title{Heat transport through lattices of quantum harmonic oscillators\\ in arbitrary dimensions}

\author{A. Asadian}
\email{ali.asadian@uibk.ac.at}
\affiliation{Institute for Theoretical Physics,
University of Innsbruck,
Technikerstr.~25, A-6020 Innsbruck, Austria}

\author{D. Manzano}
\email{daniel.manzano@uibk.ac.at}
\affiliation{Institute for Theoretical Physics,
University of Innsbruck,
Technikerstr.~25, A-6020 Innsbruck, Austria}
\affiliation{Institute for Quantum Optics and Quantum Information,
Austrian Academy of Sciences,
Technikerstr.~21A, A-6020 Innsbruck, Austria}
\affiliation{Instituto Carlos I de Fisica Te\'orica y Computational,
University of Granada, Granada, Spain}

\author{M. Tiersch}
\affiliation{Institute for Theoretical Physics,
University of Innsbruck,
Technikerstr.~25, A-6020 Innsbruck, Austria}
\affiliation{Institute for Quantum Optics and Quantum Information,
Austrian Academy of Sciences,
Technikerstr.~21A, A-6020 Innsbruck, Austria}

\author{H. J. Briegel}
\affiliation{Institute for Theoretical Physics,
University of Innsbruck,
Technikerstr.~25, A-6020 Innsbruck, Austria}
\affiliation{Institute for Quantum Optics and Quantum Information,
Austrian Academy of Sciences,
Technikerstr.~21A, A-6020 Innsbruck, Austria}

\date{\today}


\begin{abstract}
In $d$-dimensional lattices of coupled quantum harmonic oscillators, we analyze the heat current caused by two thermal baths of different temperature, which are coupled to opposite ends of the lattice, with focus on the validity of Fourier's law of heat conduction.
We provide analytical solutions of the heat current through the quantum system in the non-equilibrium steady state using the rotating-wave approximation and bath interactions described by a master equation of Lindblad form. The influence of local dephasing in the transition of ballistic to diffusive transport is investigated.
\end{abstract}

\maketitle

\section{Introduction}

Ever since the discovery of anomalous heat transport though a chain of coupled harmonic oscillators in the seminal work of Rieder, Lebowitz, and Lieb~\cite{RLL}, the topic of heat transport through systems of harmonic oscillators and the question of how classical diffusive heat transport emerges from a microscopic classical or quantum description remains an interesting field of research.
The classical law of diffusive heat conduction as stated by Fourier~\cite{Fourier},
\beq
\vec{J} = -\kappa \vec{\nabla}T,
\eeq
relates the heat current to the negative gradient of the temperature by the thermal conductivity $\kappa$, the latter being a property of the material. Fourier's law of heat conduction contains several key statements: i) at constant system size, the magnitude of the heat current is proportional to the temperature difference ($J \propto \Delta T$), ii) at constant temperature difference and constant thermal conductivity, the heat current scales inversely proportional to the distance between the heat baths, which is given by the system dimension separating the baths ($J\propto 1/L$).
Rieder, Lebowitz, and Lieb found for their one-dimensional classical system that the heat current is proportional to $\Delta T$ but does not scale with the system size. Nonetheless, when imposing Fourier's law in such situations, one generally obtains a size-dependent thermal conductivity $\kappa=\kappa(L)$, which in 1-dimensional systems usually in the form of a power-law $\kappa\propto L^\alpha$. In addition, they observed the absence of a temperature gradient \emph{inside} the system.
The same conclusions hold in higher-dimensional lattices of classical harmonic oscillators~\cite{Nakazawa}, but size-dependence has only been addressed recently, while also taking into account disorder and anharmonicity effects, e.g.\ in~\cite{RoyDhar,LeeDhar,SaitoDhar}.

In the present work, however, we focus on the \emph{quantum mechanical} equivalent of heat transport though systems of harmonic oscillators. While chains of harmonic oscillators have been investigated with an emphasis on their entanglement properties~\cite{PlenioHartleyEisert,GalveLutz}, or with the aim to scrutinize the domain of validity of master equations in Lindblad form to model thermal baths~\cite{Rivas}, specific questions regarding the heat transport properties have only been addressed to a limited extend.
Within the framework of modeling thermal baths by means of quantum Langevin equations, ballistic transport has been observed for chains of quantum harmonic oscillators~\cite{Zurcher}.
More recent analyses include the study of disordered chains, but without concrete conclusions regarding the exact scaling of the thermal conductivity with the system size~\cite{Gaul}.
Furthermore, explicit analytical results regarding the quantum mechanical steady state remain to be elucidated~\cite{DharHanggi}.
We approach this open problem by providing analytical solutions to the heat current and analytical forms of second moments of the non-equilibrium steady state in arbitrary dimensions for systems of coupled harmonic oscillators within the rotating wave approximation.

The paper is structured as follows. In the next section, we introduce the employed quantum mechanical model of a system of coupled harmonic oscillators, and the master equations describing the heat baths. We then proceed in section~\ref{solution} to the analytical solution of the heat current in the steady state for a one-dimensional chain. After commenting on the thermal nature of the state of individual oscillators in the chain in section~\ref{thermal}, we provide the analytical solution for the case that additional local dephasing influences the non-equilibrium steady state in section~\ref{dephasing}, before generalizing the result to lattices of arbitrary dimension in section~\ref{highDim}.
The results are discussed in section~\ref{discussion}, and we summarize in section~\ref{summary}.

\section{Model}

\begin{figure}
	\centering
	\includegraphics[width=0.9\linewidth]{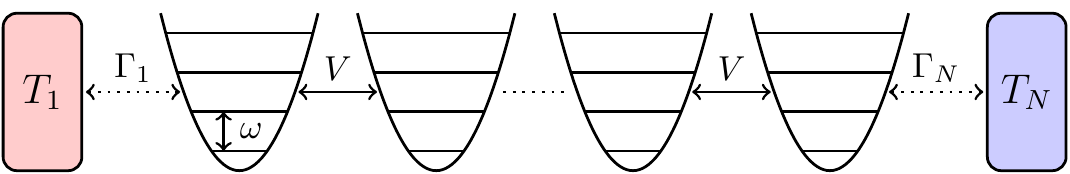}
	\caption{A linear array of coupled harmonic oscillators with thermal environments attached to the endpoints.}
	\label{cavities}
\end{figure}

The quantum system of interest is a lattice of coupled harmonic oscillators. Starting with a chain of $N$ quantum optical cavities where next-neighbors can coherently exchange photons as depicted in Fig.~\ref{cavities}, or with an array of $N$ pendula coupled by springs, one arrives after the rotating-wave approximation at an Hamiltonian of the following generic form:
\beq
\label{sysHamiltonian}
H=\sum_{j=1}^N\omega a_j^\dag a_j+\sum_{j=1}^{N-1}V_{j,j+1} \left( a_j^\dag a_{j+1}+a_{j+1}^\dag a_j \right),
\eeq
where we employ units of $\hbar=1$ throughout the paper, $\omega$ denotes the frequency of each of the $N$ identical oscillators, $V_{j,j+1}$ is the coupling strength between neighboring sites, and $a_j$ and $a_j^\dag$ are the bosonic annihilation and creation operators, respectively. The interaction term thus describes the hopping of a single excitation between neighboring sites.
Within the performed rotating wave approximation, terms proportional to $a_j a_{j+1}$ and $a_j^\dag a_{j+1}^\dag$ have been neglected. The Hamiltonian thus preserves the number of excitations and hence commutes with the excitation number operator. The rotating wave approximation is valid in the weak-coupling regime of $\omega \gg V_{j,j+1}$, which is often met in quantum optical scenarios.

The dynamics of the system described by its state $\rho$ is treated within the framework of master equations of Lindblad form, that is, the coherent transfer dynamics inside the systems on one hand, and the incoherent heat exchange with the two baths on the other:
\beq
\label{master}
\frac{d\rho}{dt}=\mathcal{L}\rho=-i[H,\rho]+\mathcal{L}_1\rho+\mathcal{L}_N\rho.
\eeq
Terms $\mathcal{L}_1 \rho$ and $\mathcal{L}_N \rho$ describe the effective, local processes of the first and last oscillator of the chain interacting with its respective heat bath. Within the Born-Markov approximation $\mathcal{L}_j\rho$ can be given in Lindblad form, and for a thermal bath it is given by
\begin{align}
\label{Lindblad}
\mathcal{L}_j\rho = &\Gamma_j(n_j+1) \left( a_j\rho a_j^\dag-\frac{1}{2}\{a_j^\dag a_j,\rho\} \right)+\\
&\Gamma_jn_j \left( a_j^\dag\rho a_j-\frac{1}{2}\{a_ja_j^\dag,\rho\} \right). \nonumber
\end{align}
The first term describes dissipation into the bath, i.e., decay of excitations into the reservoir via stimulated and spontaneous emission, respectively, and the second term describes excitation, i.e., energy absorption from the reservoir. The quantity $n_j=1/[\exp(\omega_j/k_BT_j)-1]$ denotes the mean excitation number in bath $j$ at the resonance frequency at temperature $T_j$.
This master equation of Lindblad form captures the interaction with the bosonic heat baths even at high temperatures~\cite{Rivas}.

The expression for the mean heat current through the chain can be obtained for the steady state by the following argument.
Starting with the time-derivative of the energy expectation value, which vanishes in the steady state,
\beq
\frac{d}{dt}\mean{H}=\Tr \left( H\frac{d\rho}{dt} \right)=0,
\eeq
one obtains, when inserting the master equation~\eqref{master}, that the coherent part vanishes exactly, while the contributions of the heat baths are non-zero but cancel each other. The latter amount to the positive net heat current coming from the hotter bath and the negative net heat current exiting to the colder bath, respectively:
\beq
\Tr \left(H \mathcal{L}_1\rho + H \mathcal{L}_N\rho \right) \equiv J_1 + J_N = 0.
\eeq
The net heat current through the system is thus given by $J \equiv J_1 = -J_N$.
A compact expression of $J$ for the present system employs the specific form of the Hamiltonian~\eqref{sysHamiltonian} and the terms in the master equation~\eqref{Lindblad}, and when evaluated using commutation relation $[a_j,a_j^\dag]=1$ yields:
\beq
\label{final}
J=\Gamma_1\omega_1 \Big( n_1-\mean{a_1^\dag a_1}\Big) - \frac{\Gamma_1}{2}V_{1,2} \Big(\mean{a_1^\dag a_2} + \mean{a_2^\dag a_1} \Big).
\eeq
The net heat current through the system is thus characterized by the mean excitation number of oscillator~1, which is coupled to the hotter bath, and by the coherence between this oscillator and its neighbor, oscillator~2. The heat current through the chain is therefore given by the difference between the mean energy of oscillator~1 and the equilibrium state of bath~1, and by the coherences between the two oscillators at the boundary.
An analogous expression can be derived for $J_N$, in which the corresponding terms are expressed in terms of the last oscillator, which is connected to the colder bath.

\section{Analytical solution}
\label{solution}

The coherent dynamics and the incoherent excitation exchange with the heat baths in \eqref{master} are Gaussian processes that preserve the Gaussian character of Gaussian quantum states, i.e.\ states that can be completely described by their first two moments.
Assuming that the steady state is the unique solution to $d\rho/dt = \mathcal{L}\rho = 0$ we therefore know that the steady state is Gaussian.
We thus make an ansatz for the non-equilibrium steady state using the first and second moments only.
Therefore, we first define the row vector of all bosonic operators of the system
\beq
A \equiv (a_1, a_2, \dotsc, a_N),
\eeq
and the matrix $C$, which captures the second moments, with matrix elements
\beq
[C]_{ij} \equiv \mean{a_i^\dag a_j},
\eeq
which is thus related to $A$ by taking all element-wise expectation values of the matrix $A^\dag A$.

In order to solve for the moments of the steady state, it is first necessary to transform the master equation into the equation of motion of the moments.
Since the master equation is a linear differential equation and calculating the moments is also linear in the state, we can treat the coherent and incoherent part of the master equation separately.
The coherent contribution to the equation of motion for $C$ can be obtained from the Heisenberg equation of motion for $A$.
From the time evolution of the annihilation operators,
\beq
i\dot{a}_j=[a_j,H]=\omega_j a_j+V_{j,j+1}a_{j+1}+V_{j-1,j}a_{j-1},
\eeq
in which we formally set $a_{-1}=0=a_{N+1}$, one obtains a system of coupled differential equations, the coefficients of which are contained in the matrix $W$. We thus arrive at
\beq
\dot A^\dag = i W A^\dag,
\eeq
and further, by using the product rule, at the evolution equation of $A^\dag A$.
The coherent part of the differential equation for $C$ is thus given by
\beq
\label{Ccoherent}
\left.\frac{dC}{dt}\right\vert_\text{coh.}=i [W,C].
\eeq
The effect of the baths can be calculated in a similar way by employing the incoherent part of the master equation with the exact form of $\mathcal{L}_j$ and the bosonic commutation relation. We first describe a slightly more general case where every oscillator is coupled to a local heat bath via $\mathcal{L}_j$:
\begin{align}
\label{der1}
\frac{d\mean{a_i}}{dt}&=\Tr ( a_i\mathcal{L}_i \rho) = -\frac{\Gamma_i}{2}\mean{a_i} \\
\frac{d\mean{a_i^\dag a_j}}{dt} &= \Tr ( a_i^\dag a_j[\mathcal{L}_i\rho+\mathcal{L}_j\rho]) \qquad i\ne j \nonumber\\
&= -\frac{\Gamma_i}{2} \mean{a_i^\dag a_j} -\frac{\Gamma_j}{2}\mean{a_i^\dag a_j}, \\
\frac{d\mean{a_j^\dag a_j}}{dt} &= -\Gamma_j \mean{a_j^\dag a_j} + \Gamma_j n_j.
\end{align}
The incoherent part of the evolution equation of $C$ is thus given by
\beq
\label{Cincoherent}
\left.\frac{dC}{dt}\right\vert_\text{incoh.}=\{L,C\}+M,
\eeq
where the anti-commutator is denoted by $\{L,C\}=LC+CL$, and the following diagonal matrices were introduced:
\begin{align}
L=&-\frac{1}{2}\mathrm{Diag}(\Gamma_1, \Gamma_2,\dotsc,\Gamma_N), \\
M=&\mathrm{Diag}(\Gamma_1 n_1, \Gamma_2 n_2,\dotsc,\Gamma_N n_N).
\end{align}
The complete evolution equation of $C$ for the master equation~\eqref{master} is thus given by the system of linear differential equations
\beq
\label{CMaster}
\frac{dC}{dt}=i[W,C]+\{L,C\}+M,
\eeq
where in our scenario of two heat baths at the boundaries $\Gamma_2=\dotsb=\Gamma_{N-1}=0$.
Regarding the analytic expression of the heat current~\eqref{final}, the steady state solution of $C$ contains all the information needed, and all the information necessary to describe the Gaussian steady state.

The ansatz to the solution of $C$ in the steady state, $dC_{ss}/dt=0$, consists of two contributions: one capturing the steady state of the average temperature, and one that contains the non-perturbative deviations due to the non-equilibrium.
When there is no temperature bias applied, i.e.\ $n_1=n_N \equiv n$, the equilibrium steady-state solution is given by a diagonal matrix $C=n\one$, i.e.\ with elements $[C]_{ii} = n$ on the diagonal and $[C]_{i,j\ne i}=0$ for all off-diagonal elements.
Each harmonic oscillator of the chain thus carries the same average number of excitations as both of the heat baths. This case serves as the motivation for the contribution due to the average excitation number.
In the non-equilibrium scenario, we thus choose the following ansatz
\beq
C_{ss} = \bar{n} \one + \Delta n D,
\eeq
where we have defined the average bath excitation number $\bar{n} \equiv (n_1+n_N)/2$, and the amount of non-equilibrium $\Delta n \equiv (n_1-n_N)/2$. The matrix $D$ captures the non-equilibrium contribution to the steady-state.
Insertion of this ansatz into \eqref{CMaster} yields the following equation for the non-equilibrium contribution of the steady-state:
\beq
\label{DGLC}
i[D,W] = S + \{L,D\},
\eeq
with diagonal matrix $S \equiv (2\bar{n}L+M)/\Delta n=\mathrm{Diag}(\Gamma_1,0,\dotsc,0,\Gamma_N)$.
Given the matrices $W$ with uniform couplings $V_{j,j+1}=V$, $L$, and $S$, the structure of \eqref{DGLC} implies the following form for the hermitian matrix~$D$,
\beq
\label{candi}
D = \begin{pmatrix}
e_1 & x & & &  \\
x^* & e_2 & x & &  \\
& \ddots & \ddots & \ddots & \\
& & x^* &e_{N-1} & x \\
& & & x^* & e_N
\end{pmatrix}.
\eeq
Thereby, we obtain a set of algebraic equations, which for $N>2$ reads as follows:
\begin{align*}
i\frac{V}{\Gamma_1}(x-x^*)&=1-e_1, \\ \nonumber
i\frac{V}{\Gamma_1}(e_1-e_2)&=-\frac{x}{2}, \\ \nonumber
(e_j-e_{j+1})&=0 \qquad \text{for} \quad  2\leq j\leq {N-2}, \\ \nonumber
i\frac{V}{\Gamma_N}(e_{N-1}-e_N)&=-\frac{x}{2}, \\ \nonumber
i\frac{V}{\Gamma_N}(x-x^*)&=1+e_N.
\end{align*}
The only relevant parameters involved in the solution for $D$ are the fractions $V/\Gamma_1$ and $V/\Gamma_N$. From the second and the second last equation, one can conclude that x is purely imaginary, since the $e_i$ are real numbers due to the hermiticity of~$C$. Furthermore, with the exception of $e_1$ and $e_N$, all the $e_j$ are equal. The unique solution to the above set of equations reads
\begin{align*}
x &= -i\frac{4V\Gamma_1\Gamma_N}{(4V^2+\Gamma_1\Gamma_N)(\Gamma_1+\Gamma_N)}, \\
e_1 &= \frac{4V^2(\Gamma_1-\Gamma_N)+\Gamma_1\Gamma_N^2+\Gamma_1^2\Gamma_N}{(4V^2+\Gamma_1\Gamma_N)(\Gamma_1+\Gamma_N)}, \\
e_{2\le j\le N-1}&=\frac{4V^2(\Gamma_1-\Gamma_N)+\Gamma_1\Gamma_N^2-\Gamma_1^2\Gamma_N}{(4V^2+\Gamma_1\Gamma_N)(\Gamma_1+\Gamma_N)}, \\
e_N &= \frac{4V^2(\Gamma_1-\Gamma_N)-\Gamma_1\Gamma_N^2-\Gamma_1^2\Gamma_N}{(4V^2+\Gamma_1\Gamma_N)(\Gamma_1+\Gamma_N)}.
\end{align*}
The average excitation numbers of the individual oscillators on the diagonal of $C$ is thus given by
\begin{align*}
\mean{a_1^\dag a_1} &= \bar{n}+e_1 \Delta n \\
\mean{a_j^\dag a_j} &= \bar{n}+e_j \Delta n, \qquad 2\leq j\leq {N-1} \\
\mean{a_N^\dag a_N} &= \bar{n}+e_N \Delta n,
\end{align*}
that is, the mean excitation number of all oscillators in the bulk are equal, while the excitation number of the boundary oscillators deviates from that of the bulk towards that of the heat bath. The coherences between neighboring oscillators in the diagonal above and below the main diagonal of $C$ are
\beq
\mean{a_j^\dag a_{j+1}} = x\Delta n.
\eeq

We can now give the analytic expression for the quantum heat current using the matrix elements of $C_{ss}$:
\beq
\label{analytict}
J=\frac{4\omega V^2 \Gamma_1\Gamma_N(n_1-n_N)}{(4V^2+\Gamma_1\Gamma_N)(\Gamma_1+\Gamma_N)}.
\eeq
The heat current of the chain of quantum harmonic oscillators is thus independent of the chain length~$N$, and thus constitutes a violation of Fourier's law, with the thermal conductivity scaling as $\kappa\sim N$.

As an alternative expression to \eqref{final}, we can give the heat flux in terms of the purely imaginary coherences,
\begin{equation}
\label{Jcoherence}
J=2\omega V i\mean{a_j^\dag a_{j+1}}.
\end{equation}
This means, that for the chain of oscillators of the same frequency, the heat current is solely given by the next-neighbor coherences. We thus expect any additional noise source that degrades coherent properties of the steady state to degrade the transport properties and decrease the heat current.

\section{Local thermalization}
\label{thermal}

It would be interesting to show that in steady state the local density matrices $\rho_j$ at each site are given by a thermal state of the following form when written in the Fock basis,
\begin{equation}
\rho_j=\frac{1}{\mean{n_j}+1}\sum_{m=0}^\infty \left(\frac{\mean{n_j}}{\mean{n_j}+1}\right)^m \proj{m},
\end{equation}
such that one can locally define a temperature for each oscillator in the chain.

Since the non-equilibrium steady state is of Gaussian form, the density operator can be reconstructed from the first and second moments. Each individual harmonic oscillator of the chain is thus also in a Gaussian state.
A general expression for the Gaussian density operator of a single harmonic oscillator~\cite{Adam} is given by $\rho=D(\alpha)S(r)\rho_{th}S(r)^\dag D(\alpha)^\dag$, meaning that any Gaussian state of this harmonic oscillator can be constructed from a thermal state $\rho_{th}$ by first squeezing it with $S(r,\phi)=\mathrm{exp}(\frac{1}{2}r e^{-i2\phi} a^2-\frac{1}{2}r e^{i2\phi} a^{\dag2})$, and then displacing it away from the phase space origin with $D(\alpha)=\mathrm{exp}(\alpha a^\dag-\alpha^* a)$. The mean excitation number of such a state is then given by $\mean{n}=\abs{\alpha}^2+n_{th}+(2n_{th}+1)\sinh^2r$, where $n_{th}$ is the thermal contribution to the mean excitation number.
Since due to \eqref{der1} in the steady state $\mean{a_j} = \mean{a_j^\dag}=0$, we expect $\alpha$ to vanish. Furthermore, the master equation does not involve any quadratic forms in $a$ nor $a^\dag$, and thus we also expect that the squeezing operation is not needed to describe the steady state of a single oscillator in the chain, i.e.\ the squeezing parameter $r$ should be zero. Under these conditions it is plausible that the density operator of a single oscillator at position $j$ in the chain is given by a thermal state of Gibbs form as was written above.

\section{Transition between coherent and incoherent transport}
\label{dephasing}

The ballistic transport observed for the uniform non-equilibrium chain of quantum harmonic oscillators can be turn into diffusive transport obeying Fourier's law by adding additional noise terms that degrade next-neighbor coherences. Within our master equation approach such local noise sources can be straightforwardly implemented by introducing additional dephasing terms. Local dephasing environments randomize coherences and thereby effectively degrade their magnitude. Thereby, coherently delocalized excitations are randomly localized at the sites of the harmonic oscillators, which adds a diffusive element to the transport dynamics. The Lindblad super-operator for local dephasing processes acting on each of the oscillators $j$ is given by
\begin{equation}
\mathcal{L}_{deph}\rho=\sum_{j=1}^N\gamma_j \left( a^\dag_ja_j\rho a^\dag_ja_j - \frac{1}{2} \left\{(a^\dag_ja_j)^2,\rho \right\} \right).
\end{equation}
The time evolution of the expectation value of the bosonic operators under dephasing is therefore
\begin{align*}
\frac{d\mean{a_i}}{dt} &= -\frac{\gamma_i}{2}\mean{a_i}, \\
\frac{d\mean{a^\dag_ia_j}}{dt} &= -\frac{\gamma_i}{2}\mean{a^\dag_ia_j} -\frac{\gamma_j}{2} \mean{a^\dag_ia_j} \qquad i\neq j, \nonumber \\ 
\frac{d\mean{a_i^\dag a_i}}{dt} &= 0. \nonumber
\end{align*}
With these equations, the corresponding evolution equation of $C$ in the presence of the additional dephasing environments gains additional terms:
\begin{align*}
\left.\frac{dC}{dt}\right\vert_\text{deph} =& \{L_{deph},C\} \\
& + \mathrm{Diag}\left(\gamma_1 [C]_{1,1}, \gamma_2 [C]_{2,2},\dotsc,\gamma_N[C]_{N,N}\right),
\end{align*}
where $L_{deph}=-\frac{1}{2}\mathrm{Diag}(\gamma_1,\gamma_2,\dotsc,\gamma_N)$.
Let us now consider the special case in which the local dephasing operations acting on the individual oscillators all have equal rates, i.e.\ $\gamma_j=\gamma$. This allows for a more straightforward analytical solution including the dephasing effect.
With the same ansatz as before, the non-equilibrium part of the steady-state solution satisfies the equation
\begin{align}
\label{CmasterDeph}
i[D,W]=& S+\{L,D\} -\gamma D \nonumber \\
&+\gamma\mathrm{Diag}([D]_{1,1}, [D]_{2,2},\dotsc,[D]_{N,N}).
\end{align}
The algebraic set of equations for the matrix elements of $D$ take a similar form as before, but now include the dephasing rates:
\begin{align}
i\frac{V}{\Gamma_1}(x-x^*) &= 1-e_1, \\
i\frac{V}{\Gamma_1}(e_1-e_2) &= -\frac{x}{2}-\frac{\gamma}{\Gamma_1}x, \nonumber \\
i(e_j-e_{j+1})&=\frac{\gamma}{V} x, \qquad 2\leq j\leq {N-2} \nonumber \\
i\frac{V}{\Gamma_N}(e_{N-1}-e_N) &= -\frac{x}{2}-\frac{\gamma}{\Gamma_N}x, \nonumber \\
i\frac{V}{\Gamma_N}(x-x^*) &= 1+e_N. \nonumber
\end{align}
The solution to the above set of equations are given by
\begin{align*}
x &= -i\tfrac{4V\Gamma_1\Gamma_N}{(4V^2+\Gamma_1\Gamma_N)(\Gamma_1+\Gamma_N)+2(N-1)\gamma\Gamma_1\Gamma_N}, \\ 
e_1 &= \tfrac{4V^2(\Gamma_1-\Gamma_N)+\Gamma_1\Gamma_N^2+\Gamma_1^2\Gamma_N+2(N-1)\gamma\Gamma_1\Gamma_N}{(4V^2+\Gamma_1\Gamma_N)(\Gamma_1+\Gamma_N)+2(N-1)\gamma\Gamma_1\Gamma_N}, \\
e_{2\le j\le N-1} &= \tfrac{4V^2(\Gamma_1-\Gamma_N)+\Gamma_1\Gamma_N^2-\Gamma_1^2\Gamma_N+2(N-2j+1)\gamma\Gamma_1\Gamma_N}{(4V^2+\Gamma_1\Gamma_N)(\Gamma_1+\Gamma_N)+2(N-1)\gamma\Gamma_1\Gamma_N}, \\
e_N &= \tfrac{4V^2(\Gamma_1-\Gamma_N)-\Gamma_1\Gamma_N^2-\Gamma_1^2\Gamma_N-2(N-1)\gamma\Gamma_1\Gamma_N}{(4V^2+\Gamma_1\Gamma_N)(\Gamma_1+\Gamma_N)+2(N-1)\gamma\Gamma_1\Gamma_N}.
\end{align*}
As before, this yields the unique solution to the steady state fully contained in $C_{ss}$.

In order to derive the analytical expression of the heat current in the presence of local dephasing environment, it is important to account for a possible heat current due to the dephasing environments. It is a priory not clear, that local dephasing does not introduce an additional net heat current because the local dephasing processes do not leave energy eigenstates of the chain invariant. For a chain of harmonic oscillators with uniform frequencies, and equal dephasing rates, the net heat current due to the dephasing environment, $\Tr(H\mathcal{L}_{deph}\rho)=-\frac{\gamma_1}{2}\mean{a_1^\dag a_2}-\frac{\gamma_2}{2}\mean{a_2^\dag a_1}$ vanishes for the steady state in the present scenario because coherences are purely imaginary. Thus the equality $J =J_1=-J_N$ holds under the considered dephasing operations.

With the non-equilibrium steady state solution for $C_{ss}$ the analytical expression for the heat current in the presence of dephasing environment is given by
\begin{equation}
\label{Jdeph}
J=\frac{4\omega V^2 \Gamma_1\Gamma_N(n_1-n_N)}{(4V^2+\Gamma_1\Gamma_N)(\Gamma_1+\Gamma_N)+2(N-1)\gamma\Gamma_1\Gamma_N}.
\end{equation}
The heat current with local dephasing noise on each oscillator thus acquires a dependence on the size of the system. Therefore, the heat current scales as $J \sim N^{-1}$ in the limit of large $N$, and thereby recovers a size dependence as in Fourier's law. As before, next-neighbor coherences are purely imaginary, but smaller in magnitude. The expression of the heat current in terms of coherence \eqref{Jcoherence} is also valid in the present case.

\begin{figure}
	\includegraphics[width=\linewidth]{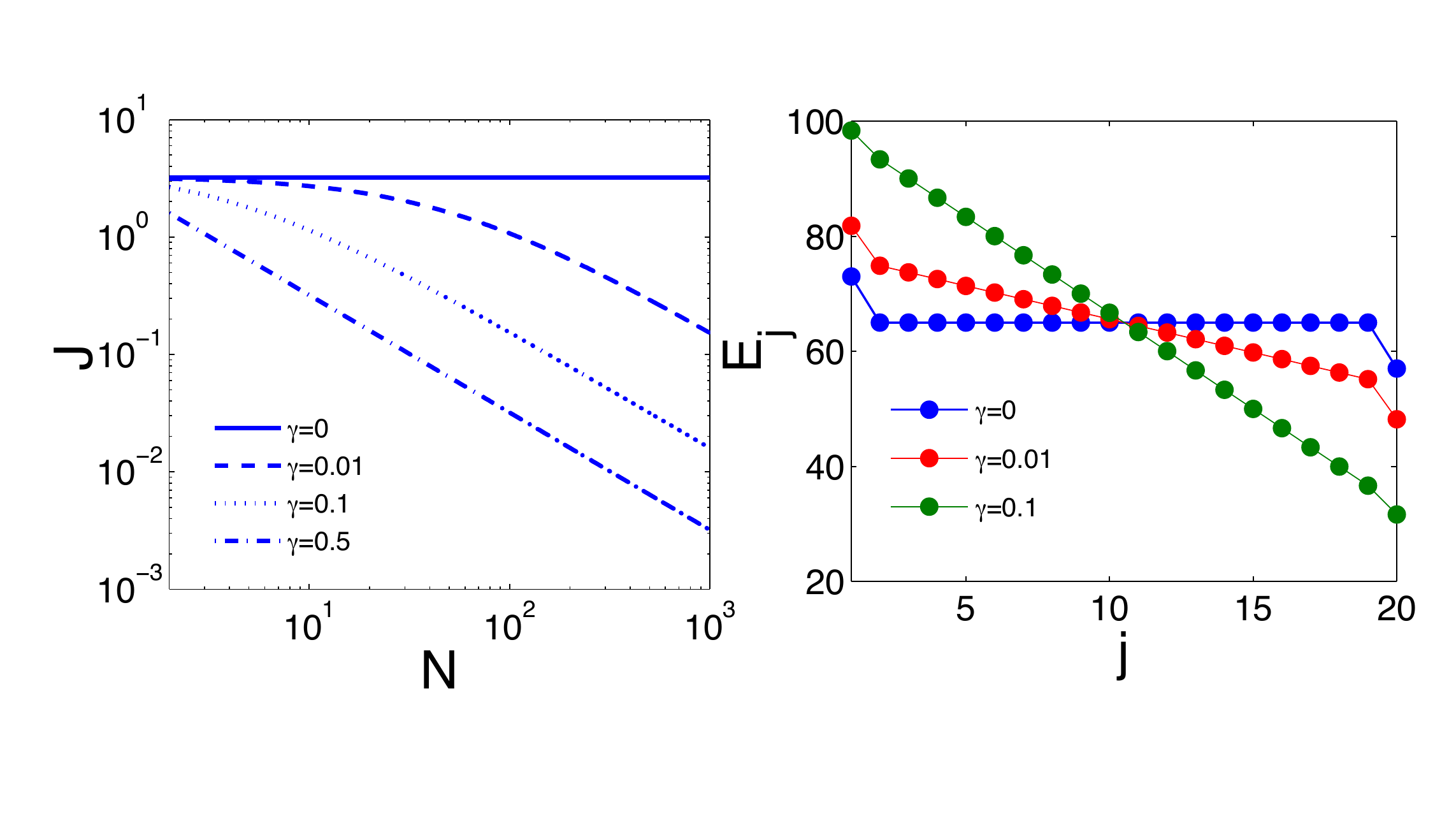}
	\caption{(Color online.)
			\textbf{Left}: Heat current vs.\ chain length in a log-log plot for different values of the dephasing rate.
			\textbf{Right}: Mean excitation number $E_j=\mean{a_j^\dag a_{j}}$ of all oscillators for a chain of length $N=20$ in the steady state under dephasing. Parameter values: $\Gamma_{1,2}=0.1$ , $V=0.1$, $\omega=10$.}
	\label{EnProfileDeph}
\end{figure}

In contrast to the transport scenario without dephasing, the mean excitations of the individual oscillators in the bulk are no longer equal, but decrease linearly from the hotter towards the colder heat bath. The effects of dephasing on the individual mean excitations of the oscillators is shown in Fig.~\ref{EnProfileDeph}. In the limit of large dephasing rates, the entire harmonic chain approaches the phenomenology of Fourier's law, in the sense that a constant gradient emerges throughout the chain.

\section{Generalization to $\boldsymbol{d}$-dimensional lattices}
\label{highDim}

\begin{figure}
	\includegraphics[width=0.9\linewidth]{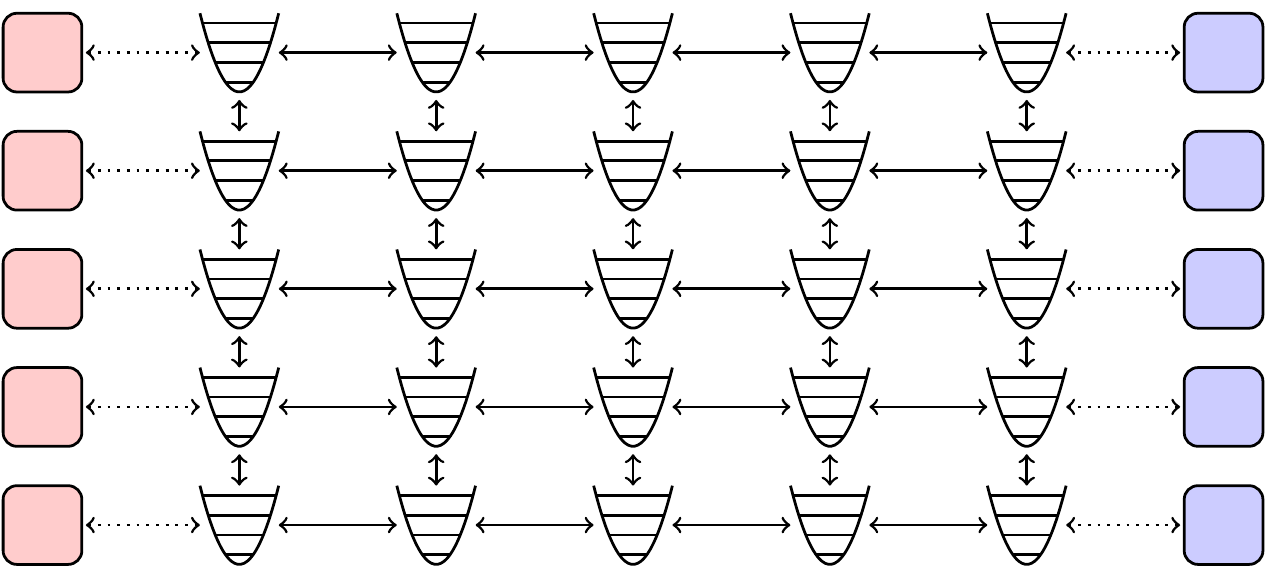}
	\caption{A 2-dimensional harmonic lattice whose boundary oscillators of opposite edges (left and right) are coupled to local heat baths of different temperature.}
	\label{2Dlattice}
\end{figure}

With the analytical results for heat transport through chains (dimension $d=1$) and the strategy to solve for steady state properties available, we generalize our results to heat transport scenarios through lattices of $d$ dimensions. The Hamiltonian~\eqref{sysHamiltonian} is extended to a cubic lattice of $d$ dimensions, again with uniform oscillator frequencies for all oscillators, and uniform couplings between next neighbors along all lattice edges.
When advancing to higher dimensions, we consider a situation where the heat transport is only driven along \emph{one} direction. Therefore, two opposing $(d-1)$-dimensional hyper-surfaces are coupled to heat baths of different temperature such that there is a temperature difference along the one remaining dimension. The two heat baths are modeled such that all harmonic oscillators of the respective hyper-surface are coupled to a local heat bath of identical temperatures.
An example, for the $d=2$ case is depicted in Fig.~\ref{2Dlattice} with the two opposing edges coupled to local heat baths.

We approach the analytic solution by relating it to the 1-dimensional case using a recursion argument. 
Clearly, due to symmetry we do not expect a net heat current perpendicular to the applied temperature gradient. However, the coherent coupling in lattice directions transverse to the temperature gradient might constructively or destructively interfere with the heat transport in the longitudinal direction.
As for the 1-dimensional case, we assume the uniqueness of the non-equilibrium steady state. This again leads us to the solution, with all essential information being captured in a matrix $C$ with matrix elements $\mean{a_{\vec{i}}^\dag a_{\vec{j}}}$, where $\vec{i}$ and $\vec{j}$ are index vectors that account for labeling of all harmonic oscillators in $d$ dimensions. We choose the order of indices such that the last index counts oscillators along the direction along which the temperature bias is applied. Thereby we can relate the heat transport through a $d$-dimensional lattice to the heat transport through $d$ stacked copies of $(d-1)$-dimensional lattices that are coherently coupled. The matrix $C$ thus is of block structure. For a $d$-dimensional hypercube of $N$ sites along every direction, $C\equiv C_d$ can be divided into $N\times N$ blocks, where each block of the diagonal captures a matrix $C_{d-1}^{(j)}$ of the $j$-th $(d-1)$-dimensional lattice, and the off-diagonal blocks capture the coherences between these different lattices of lower dimension. Since the stacked lower-dimensional lattices are only coupled to their direct neighbors, only coherences between two neighboring lattices will exists, hence only the blocks on the first diagonals above and below the main diagonal being non-zero. This procedure is repeated until the two-dimensional lattice in decomposed into an array of coupled chains.

Similarly, all the sets of differential equations can be related to those of the next-lower dimension with additional couplings along the transverse direction. We choose the same ansatz for the non-equilibrium steady state, $C_d=\bar{n}\one + \Delta n D_d$, which when inserted into the master equation yields
\beq
\label{eq:ddim}
i[D_d,W_d]=S_d + \{L_d,D_d\}.
\eeq
Let us now employ the described block structure
\begin{align*}
D_d = \begin{pmatrix}D^{(1)}_{d-1} & Q & & &  \\ Q^* & D^{(2)}_{d-1} & Q & &  \\ & \ddots & \ddots & \ddots & \\ & & Q^* &D^{(N-1)}_{d-1} & Q \\  & & & Q^* & D^{(N)}_{d-1} \end{pmatrix}, \\
W_{d} = \begin{pmatrix}W_{d-1} & G & & &  \\ G & W_{d-1} & G & &  \\ & \ddots & \ddots & \ddots & \\ & & G &W_{d-1} & G \\  & & & G & W_{d-1} \end{pmatrix},
\end{align*}
where $G=V\one$ is a diagonal matrix capturing the coupling of harmonic oscillators between two neighboring $(d-1)$-dimensional sub-lattices. The matrix $Q$ accounts for the coherences between these oscillators. The matrices $S_d$ and $L_d$ again capture the influence of the heat bath, and they are simply given by
\begin{align}
L_d=\bigoplus^{N}_{j=1}L_{d-1} \qquad \text{and} \qquad  S_d=\bigoplus^{N}_{j=1}S_{d-1},
\end{align}
with $L_1$ and $S_1$ being those for the linear chain given in Sect.~\ref{solution}.

Inserting these matrices into \eqref{eq:ddim} gives rise to the set of differential equation for the lower-dimensional sub-lattices:
\begin{align}
\label{dEq1} 
i[D^{(j)}_{d-1},W_{d-1}]&=S_{d-1}+\{L_{d-1},D^{(j)}_{d-1}\} \qquad \forall j,\\ 
\label{dEq2}
\{L_{d-1},Q_d\}&=0.
\end{align}
These equations imply that $Q_d$ vanishes and hence there are no coherences and no heat transport between sub-lattices. Furthermore, since the equations are identical for all $D^{(j)}_{d-1}$, the steady state is given by the identical steady states of the lower-dimensional sub-lattice. The non-equilibrium steady state of the $d$-dimensional lattice is thus given by the steady states of the chains connecting heat baths at different temperature to which we have decomposed the lattice.
The heat current of the $d$-dimensional lattice is thus given by the sum of all the heat currents of the identical chains composing the lattice. Hence the total heat current in $d$-dimension is the volume of the hyper-surfaces that are in contact with the heat baths, i.e.\ the number of oscillators coupled to heat baths of the same temperature, times the heat current of the chain as derived in Sec.~\ref{solution}. For the special case of a $d$-dimensional hypercube with an edge length of $N$ sites in each direction
\beq
J_d = \mathrm{Vol}_{d-1} J = N^{d-1} J.
\eeq
This implies that the heat current per chain is invariant under a length change in the direction parallel to the heat current.
The heat current of this model is ballistic also in higher dimensions.

\section{Discussion}
\label{discussion}

The analytic results for heat transport in harmonic lattices of arbitrary dimension show a number of differences to existing work. The framework of master equations in Lindblad form readily provides access to not only to steady state properties, but to the steady state itself, and it is straightforwardly applicable to many quantum optical setups and scenarios, such as the array of coupled cavities depicted in Fig.~\ref{cavities}. Our results differ from that of a harmonic chain where the baths are modeled by quantum Langevin equations~\cite{Zurcher,Gaul}. On one hand the method of treating the heat baths differs, on the other hand the ballistic transport observed in our model is independent of the system size in the absence of dephasing. Furthermore, while the mean excitation profile of the oscillators coincides in some parameter regimes, we obtain different profiles in other cases. We attribute these differences to the performed rotating wave approximation in the Hamiltonian of our model, but a detailed investigation of the effects of the rotating wave approximation on the heat transport is beyond the scope of the present work.

With respect to the open question regarding explicit results for the steady state density matrix as stated in~\cite{DharHanggi}, our work supplies analytical answers for the special case of the considered Hamiltonian and bath models, even in arbitrary dimension. The simple relation of the heat current in lattices of higher dimension may serve as a guiding result for numerical studies in higher dimensions.

Approaching the diffusive transport regime of Fourier's law by adding dephasing environments is consistent with other studies \cite{Michel,Dubi,Manzano,Znidaric}, which investigate the conditions under which the heat transport in quantum systems approaches a diffusive regime obeying Fourier's law. The mechanism applied in \cite{Michel} in order to obtain heat diffusion in quantum chains, i.e.\ introducing a band of excited states rather than sharp energy levels per site, can qualitatively be recovered by our dephasing process, i.e., the effectively introduced level fluctuations and hence the line broadening as also observed for chains of two-level atoms~\cite{Manzano}. 
The dynamic disorder introduced by the local dephasing environments allows for obtaining analytic results and a more straightforward clarification of the size-dependence than approaching the diffusive regime with disorder in the Hamiltonian as pursued in~\cite{Gaul}.

In transport scenarios of molecular biology, e.g.\ in light-harvesting systems of photosynthesis, it is being investigated to what extent entanglement or (in general non-equivalently) measures of coherence may serve as an indicator of transport properties~\cite{Sarovar,Fassioli,Scholak,Ishizaki,Tiersch}.
While we have observed the emergence of entanglement in strongly non-equilibrium systems of coupled two-level atoms~\cite{Manzano}, it does not hold for systems of harmonic oscillators.

This result is not surprising given that operators of the form $a^2$ and $(a^\dag)^2$ have been neglected in the Hamiltonian and the master equation, which are required, however, to achieve a squeezing operation that is necessary for creating an entangled Gaussian state. The conclusion to be drawn is that although the system can exhibit large coherences and thereby a large heat current, i.e.\ efficient transport, entanglement is completely absent and also cannot be generated by a large temperature difference as possible for systems of two-level atoms~\cite{Manzano}.
As spelled out by~\eqref{Jcoherence}, the magnitude of the heat current depends on the amount of quantum coherence between the neighboring oscillators. Therefore, entanglement certainly does not qualify as a signature of quantum transport efficiency in the present model.

The validity of Fourier's law in higher dimensions strongly depends on the modeling of the system. Most of the works in this direction highly rely on large-scale simulations~\cite{LeeDhar,SaitoDhar,Yang}.
Peierls showed that the phonon-phonon scattering leads to diffusive heat conduction in higher dimensional systems~\cite{Peierls}.
In~\cite{Lepri} it is shown, that all models characterized by short range interactions and momentum conservation should exhibit the same kind of anomalous behavior in the heat conduction, i.e.\ a divergence of the thermal conductivity at infinite system-size, for $d<3$.
We can thus extend the result of~\cite{Lepri} for the specific case of harmonic oscillator lattices to arbitrary dimensions~$d$ with an anomalous heat conduction with $\kappa \propto N$.

\section{Summary}
\label{summary}

We provide an analytical solution to the heat current and the non-equilibrium steady state of a chain of quantum harmonic oscillators, whose boundary oscillators are coupled to two heat baths of different temperature, respectively. The heat current is independent of the chain length, and thus is of ballistic nature exhibiting an anomalous heat current with a thermal conductivity proportional to the chain length. The diffusive heat transport regime, i.e.\ normal heat conduction with a constant thermal conductivity, is recovered when additional dephasing environments locally affect each of the oscillators. We observe the absence of entanglement in the chain for all parameter regimes. Finally, we provide the analytical expression for the heat current of a $d$-dimensional lattice of harmonic oscillators, which turns out to be the sum of the heat currents of all the chains, into which the lattice may be decomposed, each one connecting the two heat baths.

\begin{acknowledgments}
The research was funded by the Austrian Science Fund (FWF): F04011 and F04012.
D.M. acknowledges funding from Spanish MEC-FEDER, project FIS2009-08451, together with the Campus de Excelencia Internacional and the Junta de Andalucia, project FQM-165.
\end{acknowledgments}



\end{document}